\documentclass[twocolumn,showpacs,amsmath,amssymb,superscriptaddress]{revtex4-1}

\usepackage{dcolumn}

\usepackage{graphicx}
\usepackage{dcolumn}
\usepackage{bm}

\begin{document}


\title{Quadrupole Collective Dynamics from Energy Density Functionals: 
Collective Hamiltonian and the Interacting Boson Model}

\author{K.~Nomura}
\affiliation{Department of physics, University of Tokyo, Hongo,
Bunkyo-ku, Tokyo, 113-0033, Japan} 
\affiliation{Physics Department, Faculty of Science, University of
Zagreb, 10000 Zagreb, Croatia}

\author{T.~Nik$\check{\rm s}$i\'c}
\affiliation{Physics Department, Faculty of Science, University of
Zagreb, 10000 Zagreb, Croatia} 

\author{T.~Otsuka}
\affiliation{Department of physics, University of Tokyo, Hongo,
Bunkyo-ku, Tokyo, 113-0033, Japan} 
\affiliation{Center for Nuclear Study, University of Tokyo, Hongo,
Bunkyo-ku Tokyo, 113-0033, Japan} 
\affiliation{National Superconducting Cyclotron Laboratory, 
Michigan State University, East Lansing, MI}

\author{N.~Shimizu}
\affiliation{Department of physics, University of Tokyo, Hongo,
Bunkyo-ku, Tokyo, 113-0033, Japan} 
\affiliation{Center for Nuclear Study, University of Tokyo, Hongo,
Bunkyo-ku Tokyo, 113-0033, Japan} 

\author{D.~Vretenar}
\affiliation{Department of physics, University of Tokyo, Hongo,
Bunkyo-ku, Tokyo, 113-0033, Japan} 
\affiliation{Physics Department, Faculty of Science, University of
Zagreb, 10000 Zagreb, Croatia}

\date{\today}

\begin{abstract}

Microscopic energy density functionals (EDF) have become a standard tool
for nuclear structure calculations, providing an accurate global
description of nuclear ground states and collective excitations. 
For spectroscopic applications 
this framework has to be extended to account for collective correlations 
related to restoration of symmetries broken by the static mean field, and 
for fluctuations of collective variables. In this work we compare two approaches 
to five-dimensional quadrupole dynamics: the collective Hamiltonian for 
quadrupole vibrations and rotations, and the Interacting Boson Model.
The two models are compared in a study of the evolution of non-axial shapes in 
Pt isotopes. Starting from the binding energy surfaces of $^{192,194,196}$Pt, 
calculated with a microscopic energy density functional, we analyze 
the resulting low-energy collective spectra obtained from the collective Hamiltonian, 
and the corresponding IBM-2 Hamiltonian.  The calculated excitation spectra and 
transition probabilities for the ground-state bands and the $\gamma$-vibration bands 
are compared to the corresponding sequences of experimental states.
\end{abstract}

\pacs{21.10.Re,21.60.Ev,21.60.Fw,21.60.Jz}

\maketitle



\section{Introduction}

One of the major research topics in the theoretical nuclear structure physics 
has been the study of quadrupole collective dynamics from a
microscopic viewpoint \cite{BM,RS,Bender_review,QPT_review}. 
Quadrupole collectivity results from multi-nucleon dynamics of
nuclear surface deformation. The equilibrium shape of a nucleus 
can change depending on the number of valence 
nucleons: a spherical vibrator, a deformed rotor, or a soft 
shape in between. In most isotopic or isotonic sequences the
transition between different shapes is gradual, but in a number of
cases, with the addition or subtraction of only few nucleons, one finds signatures 
of abrupt changes in observables that characterize equilibrium shapes. 
These structure phenomena have been investigated using concepts of 
quantum shape/phase transitions \cite{QPT_review}, and advanced 
self-consistent beyond-mean-field 
approaches \cite{Rayner2010SrZrMo,Rayner2010EFA,Ni.07,Li2009NdSmGd,Li2010BaXe}.  

Microscopic studies based on energy density functionals (EDFs) have been quite
successful in reproducing with remarkable accuracy various intrinsic
(bulk) properties of medium-mass and heavy nuclei such as binding energies, 
density distributions, charge radii, giant resonances, etc \cite{RS,Bender_review}. 
The current generation of EDFs includes non-relativistic Skyrme- \cite{Sk,VB} and
Gogny- \cite{Go,D1S} functionals, as well as relativistic
density functionals \cite{Vretenar-1,RMF_review}. 
The framework of EDFs has also been extended beyond the mean-field level to describe 
excitation spectra and electromagnetic transition rates. Models have been developed 
that perform restoration of symmetries broken by the static nuclear mean field, and take into account 
quadrupole fluctuations: GCM configuration mixing calculations  
\cite{RS,Bender_review,Rayner2002GCMMg,TRodriguez2010triaxial,Yao2011GCM2},
and solutions of the collective Hamiltonian 
with quadrupole degrees of freedom \cite{CollSk,CollGo,Nik2009Gd,Li2009NdSmGd,Li2010BaXe}. 

Another successful approach to the low-lying structure of medium-mass and
heavy nuclei is based on the interacting boson approximation \cite{IBM}. 
The Interacting Boson Model (IBM), in particular,  provides not only an algebraic but 
in some cases also a microscopic description of 
nearly-spherical and $\gamma$-unstable shapes \cite{IBM,OAI,OCas,MO,nso,nsofull}. 
From a microscopic point of view, the collective $J=0^{+}$ and $2^{+}$
pairs of valence protons (neutrons) 
are mapped onto the corresponding boson images with $J=0^{+}$ and $2^{+}$, denoted by
$s_{\pi}$ ($s_{\nu}$) and $d_{\pi}$ ($d_{\nu}$) bosons, respectively \cite{OAI}. 
A number of studies have been carried out to derive the IBM Hamiltonian starting from
nucleonic degrees of freedom in terms of the conventional mapping method that starts
from the shell model \cite{OAI}, the more recent approach based on binding energy maps 
calculated with microscopic EDFs \cite{nso}, etc.

A static self-consistent mean-field solution in the intrinsic frame, for instance a 
map of the energy surface as a function of quadrupole deformation, is characterized by 
symmetry breaking: translational, rotational, particle number, and can only
provide an approximate description of bulk ground-state properties.
To calculate excitation spectra and electromagnetic transition rates in individual nuclei, it is 
necessary to include correlations that arise from symmetry restoration and fluctuations 
around the mean-field minimum. Both types of correlations can be included simultaneously by 
mixing angular-momentum projected states corresponding to different quadrupole moments. 
The most effective approach for configuration mixing
calculations is the generator coordinate method (GCM), 
with multipole moments used as coordinates that
generate the intrinsic wave functions. It must be noted that, while GCM configuration
mixing of axially symmetric states has been implemented by several groups and 
routinely used in nuclear structure studies
\cite{Bender2004Pb,Nik2006GCM2axial,TRodriguez2008NdPLB}, 
the application of this method to triaxial 
shapes presents a much more involved and technically difficult problem
\cite{Bender2008GCMtriaxial,Yao2011GCM2}. 
In addition, the use of general EDFs, that is, with
an arbitrary dependence on nucleon densities, in GCM type
calculations, often leads to discontinuities or even divergences of
the energy kernels as a function of deformation~\cite{AER.01,Dob.07}.
Only for certain types of density dependence a regularization method
can be implemented, which corrects energy kernels and
removes the discontinuities and divergences 
\cite{SkyrmeGCM2009_1,SkyrmeGCM2009_2,SkyrmeGCM2009_3}. 

In an approximation to the full GCM approach to five-dimensional quadrupole dynamics that 
restores rotational symmetry and allows for fluctuations around the triaxial mean-field minima, 
a collective Hamiltonian can be formulated, with deformation-dependent parameters determined 
by constrained microscopic self-consistent mean-field calculations. The dynamics of the five-dimensional 
Hamiltonian for quadrupole vibrational and rotational degrees of freedom is governed by the seven functions 
of the intrinsic quadrupole deformations: the collective potential, three vibrational mass parameters, and 
three moments of inertia for rotations around the principal axes 
\cite{CollSk,CollGo,Nik2009Gd,Li2009NdSmGd,Li2010BaXe}.

Another approximation consists in mapping the self-consistent mean-field solution 
to a boson (IBM) Hamiltonian. In Refs.~\cite{nso,nsofull} the energy surface for
quadrupole degrees of freedom,  calculated from a microscopic EDF, was 
mapped onto the corresponding boson energy surface under certain approximations. 
The interaction strengths of the boson Hamiltonian are determined by the mapping procedure. 
One then proceeds to calculate the excitation spectra and wave functions in the laboratory 
frame \cite{nso,nsofull}.  The validity of the method of Ref.~\cite{nso} was tested in various 
mass regions \cite{nsofull,GognyIBMPt,osw-gogny}. 

It would be, therefore, interesting to compare the two approximations starting 
from the same self-consistent mean-field solution based on a microscopic 
EDF. In this work we compare spectroscopic observables calculated with 
the IBM Hamiltonian to the solution of the collective quadrupole Hamiltonian, 
with both calculations based on relativistic Hartree-Bogoliubov (RHB) \cite{Vretenar-1}
self-consistent binding energy surfaces. The framework of relativistic EDFs 
and the corresponding collective Hamiltonian have successfully been employed 
in studies of the evolution of ground-state shapes and spectroscopic properties of 
medium-heavy and heavy nuclei~\cite{Nik2009Gd,Li2009NdSmGd,Li2010BaXe,RMF_review,RMFPt}.  
In the present analysis we consider the even-even isotopes $^{192-196}$Pt. In the 
IBM framework these $\gamma$-soft nuclei can be characterized by the 
O(6) dynamical symmetry \cite{IBM,Pt196O6,CastenCizewski}. 

The paper is organized as follows. 
In Sec.~\ref{sec:th}, we briefly describe the theoretical procedures used 
to derive the collective Hamiltonian and the IBM Hamiltonian starting 
from a given EDF. The microscopic RHB energy surface and
the mapped IBM energy surface are discussed in Sec.~\ref{sec:pes}. 
Spectroscopic properties of $^{192-196}$Pt calculated with the two 
models are compared in Sec.~\ref{sec:spectra}. 
Section~\ref{sec:summary} summarizes the results and presents a 
short outline of future work. 

\section{Theoretical framework}
\label{sec:th}

The map of the energy surface as a function of the
quadrupole collective variables $\beta$ and $\gamma$ \cite{BM} 
is obtained from self-consistent RHB calculations with additional 
constraints on the axial and triaxial mass quadrupole moments. 
The quadrupole moments can be related to the polar
deformation parameters $\beta$ and $\gamma$. 
The parameter $\beta$ is simply proportional to the intrinsic quadrupole moment,
and the angular variable $\gamma$ specifies the type and orientation of the shape. 
The limit $\gamma = 0$ corresponds to axial
prolate shapes, whereas the shape is oblate for $\gamma = \pi/3$. 
Triaxial shapes are associated with intermediate
values $0 < \gamma < \pi/3$. In this work the constrained RHB calculations have been 
performed using the relativistic functional DD-PC1 \cite{DD-PC1}. Starting from
microscopic nucleon self-energies in nuclear matter, and empirical global properties of the nuclear matter
equation of state, the coupling parameters of DD-PC1 have been determined in a careful comparison of the
calculated binding energies with data, for a set of 64 axially deformed nuclei 
in the mass regions $A \approx 150 - 180$
and $A \approx 230 - 250$. DD-PC1 has been further tested in a series of calculations of properties of spherical and
deformed medium-heavy and heavy nuclei, including binding energies, charge radii, deformation parameters,
neutron skin thickness, and excitation energies of giant multipole resonances. For the examples presented
here, pairing correlations have been taken into account by employing a pairing force that is separable in
momentum space, and is completely determined by two parameters adjusted to reproduce the empirical 
bell-shaped pairing gap in symmetric nuclear matter \cite{pairing}.

The entire dynamics of the collective Hamiltonian is governed by 
seven functions of the intrinsic deformations $\beta$ and $\gamma$:
the collective potential, the three mass parameters:
$B_{\beta\beta}$, $B_{\beta\gamma}$, $B_{\gamma\gamma}$, and the
three moments of inertia $\mathcal{I}_k$. These functions are
determined by the choice of a particular microscopic nuclear energy
density functional and a pairing functional. 
The quasiparticle wave functions and energies, that correspond to  
constrained self-consistent solutions of the RHB model, 
provide the microscopic input for the parameters of the
collective Hamiltonian \cite{Nik2009Gd}: 
\begin{equation}
\label{hamiltonian-quant}
\hat{H}_{\textnormal{coll}} = \hat{T}_{\textnormal{vib}}+\hat{T}_{\textnormal{rot}}
              +V_{\textnormal{coll}} \; ,
\end{equation}
with the vibrational kinetic energy:
\begin{eqnarray}
\hat{T}_{\textnormal{vib}} =
&-&\frac{\hbar^2}{2\sqrt{wr}}
   \Big\{\frac{1}{\beta^4}
   \Big[\frac{\partial}{\partial\beta}\sqrt{\frac{r}{w}}\beta^4
   B_{\gamma\gamma} \frac{\partial}{\partial\beta}
   \nonumber \\
   &-&\frac{\partial}{\partial\beta}\sqrt{\frac{r}{w}}\beta^3
   B_{\beta\gamma}\frac{\partial}{\partial\gamma}
   \Big]
   +\frac{1}{\beta\sin{3\gamma}}\Big[
   -\frac{\partial}{\partial\gamma} \sqrt{\frac{r}{w}}\sin{3\gamma}
\nonumber \\
      &\times&
B_{\beta \gamma}\frac{\partial}{\partial\beta}
    +\frac{1}{\beta}\frac{\partial}{\partial\gamma} \sqrt{\frac{r}{w}}\sin{3\gamma}
      B_{\beta \beta}\frac{\partial}{\partial\gamma}
   \Big]\Big\} \; ,
\end{eqnarray}
and rotational kinetic energy:
\begin{equation}
\hat{T}_{\textnormal{\textnormal{\textnormal{rot}}}} =
\frac{1}{2}\sum_{k=1}^3{\frac{\hat{J}^2_k}{\mathcal{I}_k}} \; .
\end{equation}
$V_{\textnormal{coll}}$ is the collective potential.
$\hat{J}_k$ denotes the components of the angular momentum in
the body-fixed frame of a nucleus, and the mass parameters
$B_{\beta\beta}$, $B_{\beta\gamma}$, $B_{\gamma\gamma}$, as well as
the moments of inertia $\mathcal{I}_k$, depend on the quadrupole
deformation variables $\beta$ and $\gamma$:
$\mathcal{I}_k = 4B_k\beta^2\sin^2(\gamma-2k\pi/3)$.
Two additional quantities that appear in the expression for the vibrational energy:
$r=B_1B_2B_3$, and $w=B_{\beta\beta}B_{\gamma\gamma}-B_{\beta\gamma}^2 $,
determine the volume element in the collective space. The moments of inertia are
computed using the Inglis-Belyaev (IB) formula \cite{Inglis,Belyaev}, and the mass
parameters associated with the two quadrupole collective coordinates
$q_0=\langle\hat{Q}_{20}\rangle$ and $q_2=\langle\hat{Q}_{22}\rangle$
are calculated in the cranking approximation. The potential $V_{\textnormal{coll}}$ 
in the collective Hamiltonian
Eq.~(\ref{hamiltonian-quant}) is obtained by subtracting the zero-point
energy corrections from the total energy that corresponds to the
solution of constrained RHB equations, at each point on the triaxial
deformation plane. The Hamiltonian Eq.~(\ref{hamiltonian-quant}) describes quadrupole vibrations,
rotations, and the coupling of these collective modes. The corresponding
eigenvalue problem is solved using an expansion of eigenfunctions in terms
of a complete set of basis functions that depend on the deformation variables $\beta$ and
$\gamma$, and the Euler angles $\phi$, $\theta$ and $\psi$ \cite{Nik2009Gd}. 
The diagonalization of the Hamiltonian yields the excitation energies and collective
wave functions for each value of the total angular momentum and parity, 
that are used to calculate observables. An important advantage of using
the collective model based on self-consistent mean-field single-(quasi)particle solutions is the fact 
that physical observables, such as transition probabilities and spectroscopic quadrupole 
moments, are calculated in the full configuration space and
there is no need for effective charges. Using the bare value of the proton charge in the
electric quadrupole operator, the transition probabilities between eigenvectors of
the collective Hamiltonian can be directly compared with data.

In an equivalent approach the RHB binding energy surface can be mapped onto the IBM 
Hamiltonian. Starting from the energy surface $E_{\rm RMF}(\beta,\gamma)$ calculated 
with the DD-PC1 plus separable-pairing functional, each point on the 
$(\beta,\gamma)$ plane is 
mapped onto the corresponding point on the energy surface calculated in
the IBM, referred to hereafter as $E_{\rm IBM}(\beta_{B},\gamma_{B})$,
using the method proposed in Ref.~\cite{nsofull}. 
Here $\beta_{B}$ and $\gamma_{B}$ denote the boson images of 
the quadrupole deformation parameters $\beta$ and
$\gamma$, respectively, that are used as constraints in the self-consistent RHB 
calculation and appear as variables in the collective Hamiltonian. 
The boson images $\beta_{B}$ and $\gamma_{B}$ are related to $\beta$ and $\gamma$ through
the proportionality $\beta_{B}\propto\beta$, and the equality
$\gamma_{B}=\gamma$, respectively \cite{nso,nsofull}. 
This mapping procedure is used to determine the strength parameters of the IBM Hamiltonian. 

We consider the IBM-2 model \cite{OAI}: the number of proton (neutron) bosons, 
denoted by $n_{\pi}$ ($n_{\nu}$),
are assumed to equal half the number of valence protons
(neutrons). In the consistent-Q formalism \cite{IBM}
the IBM-2 Hamiltonian reads: 
\begin{eqnarray}
 \hat H_{\rm IBM}=\epsilon(\hat n_{d\pi}+\hat n_{d\nu})+\kappa\hat
  Q_{\pi}\cdot\hat Q_{\nu}, 
\label{eq:bh}
\end{eqnarray}
where $\hat n_{d\rho}=d_{\rho}^{\dagger}\cdot\tilde d_{\rho}$ ($\rho=\pi$ or $\nu$) and 
$\hat Q_{\rho}=[s_{\rho}^{\dagger}\tilde d_{\rho}+d_{\rho}^{\dagger}\tilde s_{\rho}]^{(2)}+\chi_{\rho}[d_{\rho}^{\dagger}\tilde d_{\rho}]^{(2)}$ denote the $d$-boson number-operator and the quadrupole operator, respectively. 
$\epsilon$ and $\kappa$ are coupling constants. 
The parameters $\chi_{\pi,\nu}$ inside the quadrupole operators are
quite relevant to determining whether a nucleus is prolate or oblate
deformed.

The bosonic energy surface $E_{\rm IBM}(\beta,\gamma)$ corresponds 
to the classical limit of the Hamiltonian $\hat H_{\rm IBM}$:
$E_{\rm IBM}(\beta_{B},\gamma_{B})=\langle\Psi(\beta_{B},\gamma_{B})|\hat H_{\rm
IBM}|\Psi(\beta_{B},\gamma_{B})\rangle$.
$|\Psi(\beta_{B},\gamma_{B})\rangle$ denotes the boson coherent state \cite{coherent}:
$|\Psi(\beta_{B},\gamma_{B})\rangle\propto\prod_{\rho=\pi,\nu}[s_{\rho}^{\dagger}+\beta_{\rho}\cos{\gamma_{\rho}}d_{\rho
0}^{\dagger}+\frac{1}{\sqrt{2}}\beta_{\rho}\sin{\gamma_{\rho}}(d_{\rho +2}^{\dagger}+d_{\rho -2}^{\dagger})]^{n_{\rho}}|0\rangle$, up to a normalization constant. 
Here $|0\rangle$ is the boson vacuum, and the variables $\beta_{\rho}$ and
$\gamma_{\rho}$ are the corresponding polar deformation parameters. 
As in our previous studies \cite{nso,nsofull}, it is assumed that $\beta_{\pi}=\beta_{\nu}\equiv\beta_{B}$
and $\gamma_{\pi}=\gamma_{\nu}\equiv\gamma_{B}$. 
The analytical form of $E_{\rm IBM}(\beta_{B},\gamma_{B})$ can be found in
Refs.~\cite{nso,nsofull}. Hereafter we denote  the bosonic energy surface as 
$E_{\rm IBM}(\beta,\gamma)$, omitting the indices of $\beta_{B}$ and
$\gamma_{B}$. 

The boson Hamiltonian $\hat H_{\rm IBM}$, parametrized by the microscopically 
calculated coupling constants, is diagonalized in the $M=0$ boson space. 
Here $M$ denotes the $z$-component of the total boson angular momentum $L$. 
Reduced quadrupole transition probabilities $B$(E2) are calculated for transitions between 
the eigenstates of the IBM Hamiltonian. 

Here we point out again that the total boson energy $E_{\rm IBM}(\beta,\gamma)$ 
has been related to the microscopic EDF energy surface (total energy). 
However, for the IBM Hamiltonian $\hat H_{\rm IBM}$ one cannot make a
distinction between the kinetic and potential terms, as in the
corresponding collective Hamiltonian $\hat H_{\rm coll}$.  
Nevertheless, the effects relevant to both vibrational and rotational
kinetic energies are assumed to be incorporated into the IBM approach by adjusting 
$E_{\rm IBM}(\beta,\gamma)$ to be as close as possible to the 
microscopic surface $E_{\rm RHB}(\beta,\gamma)$.  
This prescription turned out to be valid for vibrational and 
$\gamma$-soft nuclei at moderate quadrupole deformation
\cite{nso,nsofull}, similarly to the
conventional mapping method of Ref.~\cite{OAI}. 
For rotational nuclei with large quadrupole deformation,
however, the overall scale of the IBM rotational spectra 
differs from the experimental one \cite{nso,nsofull}. 
The discrepancy partially arises because nuclear rotational properties,
characterized by the overlap of the intrinsic state and
the rotated one, differs from the rotational characteristics 
of the corresponding boson system \cite{IBMrot}. 
This problem may be cured by the recently proposed prescription
\cite{IBMrot}, in which the rotational response (i.e., cranking) of 
boson system is related to the rotational response of nucleon system. 
This procedure goes beyond the simple analysis of the zero-frequency 
energy surface.  In order that the boson rotational response becomes 
equal to the fermion (nucleon) response,  an additional kinetic term
$\hat L\cdot\hat L$  has to be included in the boson Hamiltonian, with a 
coupling constant determined microscopically \cite{IBMrot}. 
The term $\hat L\cdot\hat L$ directly influences 
the moment of inertia of rotational band with the eigenvalue $L(L+1)$. 
However, the above-mentioned problem, concerning the IBM rotational
spectra, does not occur in the considered Pt nuclei, 
and thus one does not need to include the $\hat L\cdot\hat L$ term in
the present case.   

Similar problems with the overall scale of the rotational spectra are 
also encountered in the collective Hamiltonian model, when the 
IB formula is used to calculate 
the moments of inertia \cite{Nik2009Gd,Li2009NdSmGd,Li2010BaXe}. 
The inclusion of an additional scale parameter is often necessary because of the well known
fact that the IB formula predicts effective moments of inertia that are
considerably smaller than empirical values. More realistic values are only obtained if one
uses the Thouless-Valatin (TV) formula, but this procedure is computationally much more
demanding. In the present case we have used the IB moments of inertia in the 
calculation of excitation spectra of Pt nuclei, and the agreement with experiment 
is such that no renormalization of the effective moments of inertia is required. 
This result allows for a direct comparison of the IBM spectra to 
the solutions of the collective Hamiltonian.

\section{Binding energy surfaces in the $\beta - \gamma$ plane}
\label{sec:pes}

\begin{figure*}[ctb!]
\begin{center}
\begin{tabular}{cc}
\includegraphics[width=7.0cm]{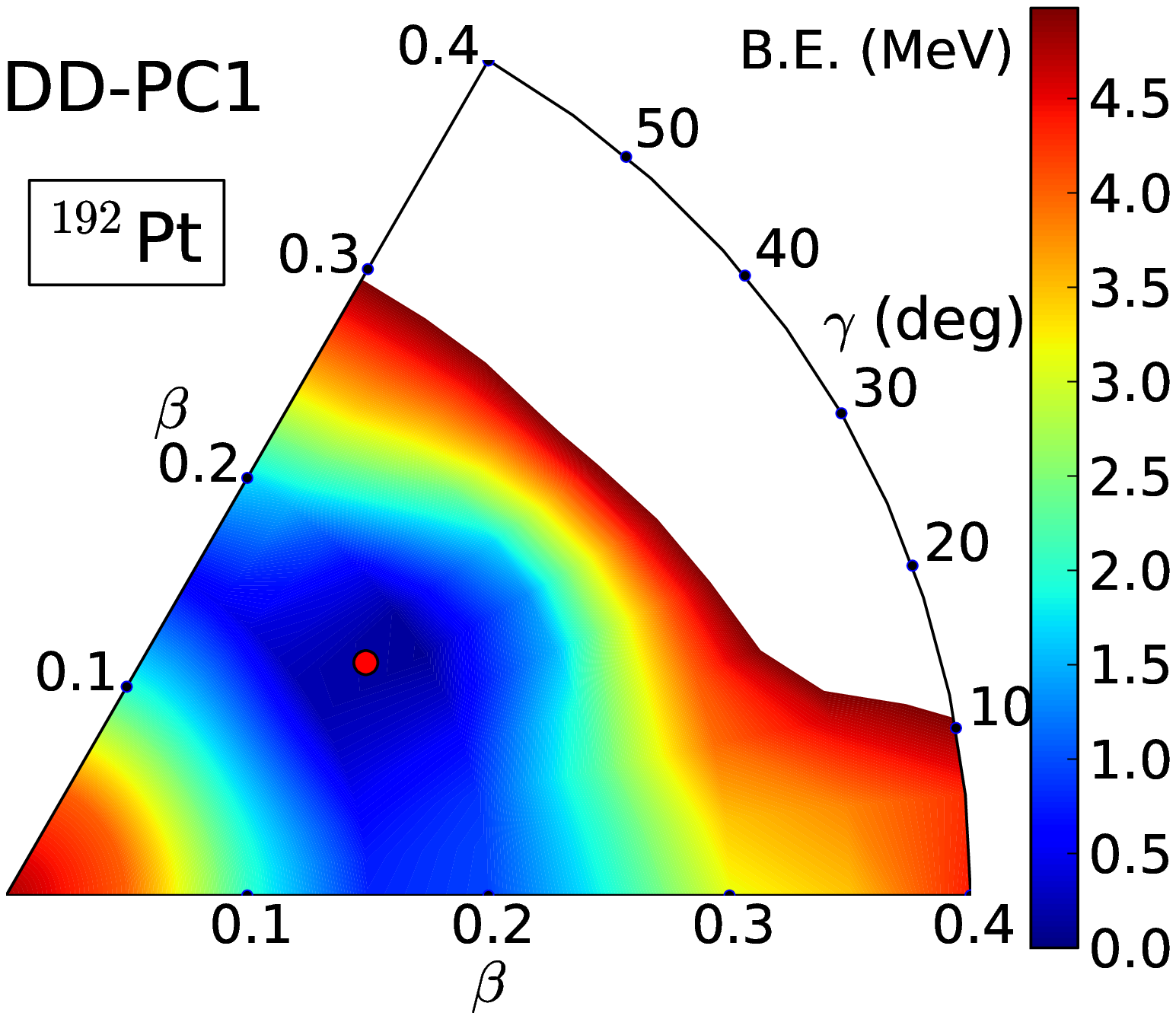} &
\includegraphics[width=7.0cm]{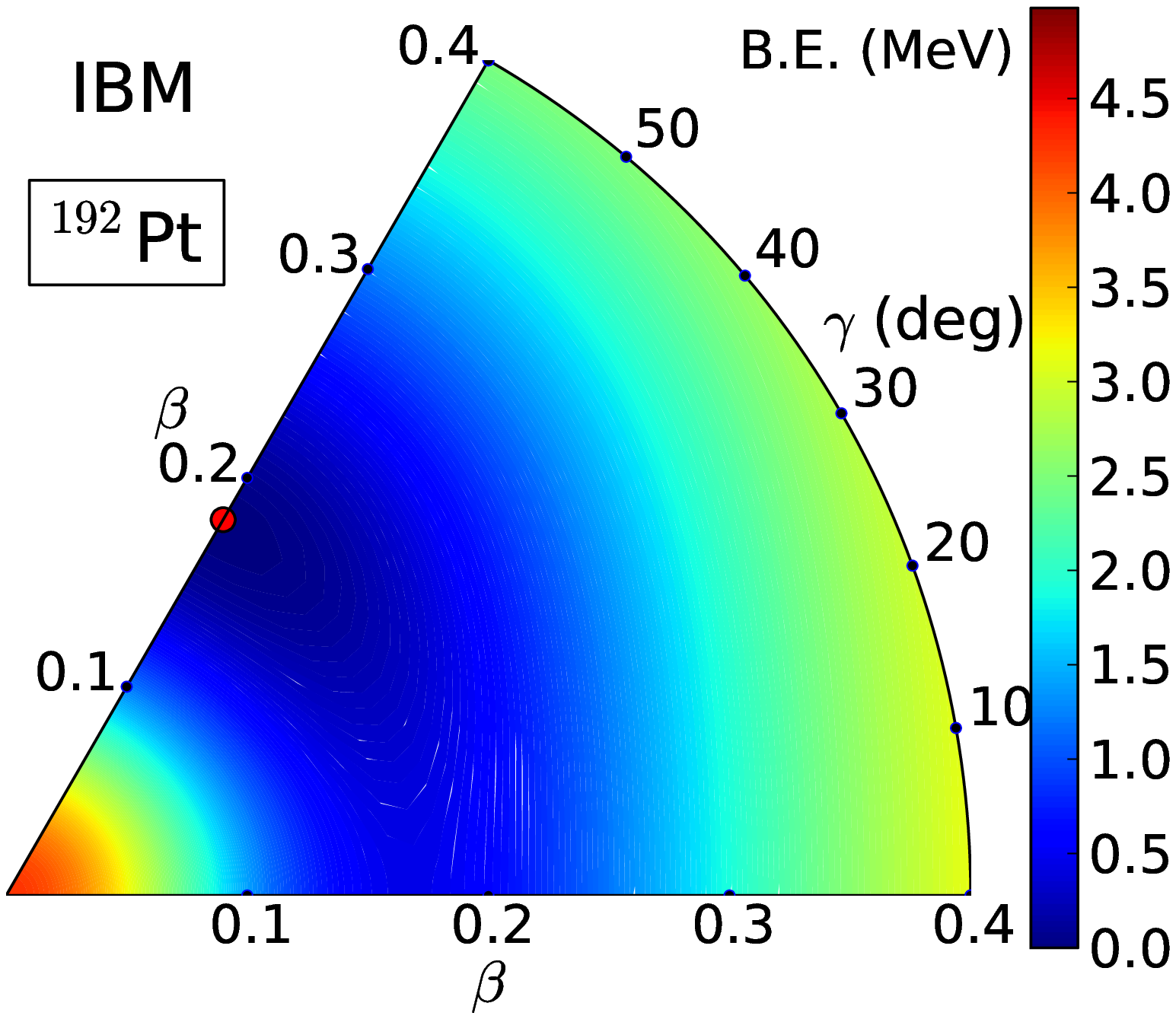}\\
\includegraphics[width=7.0cm]{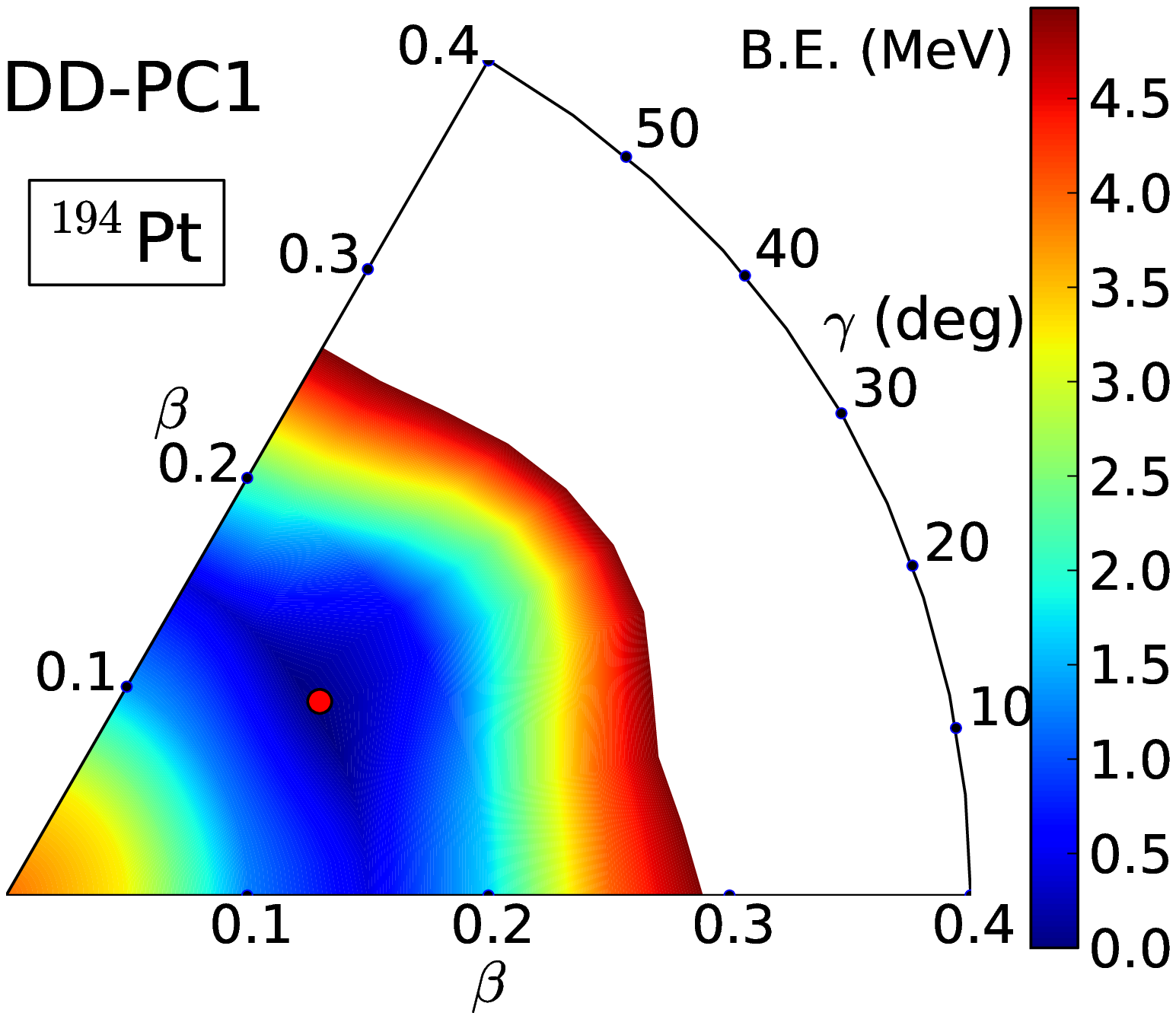} &
\includegraphics[width=7.0cm]{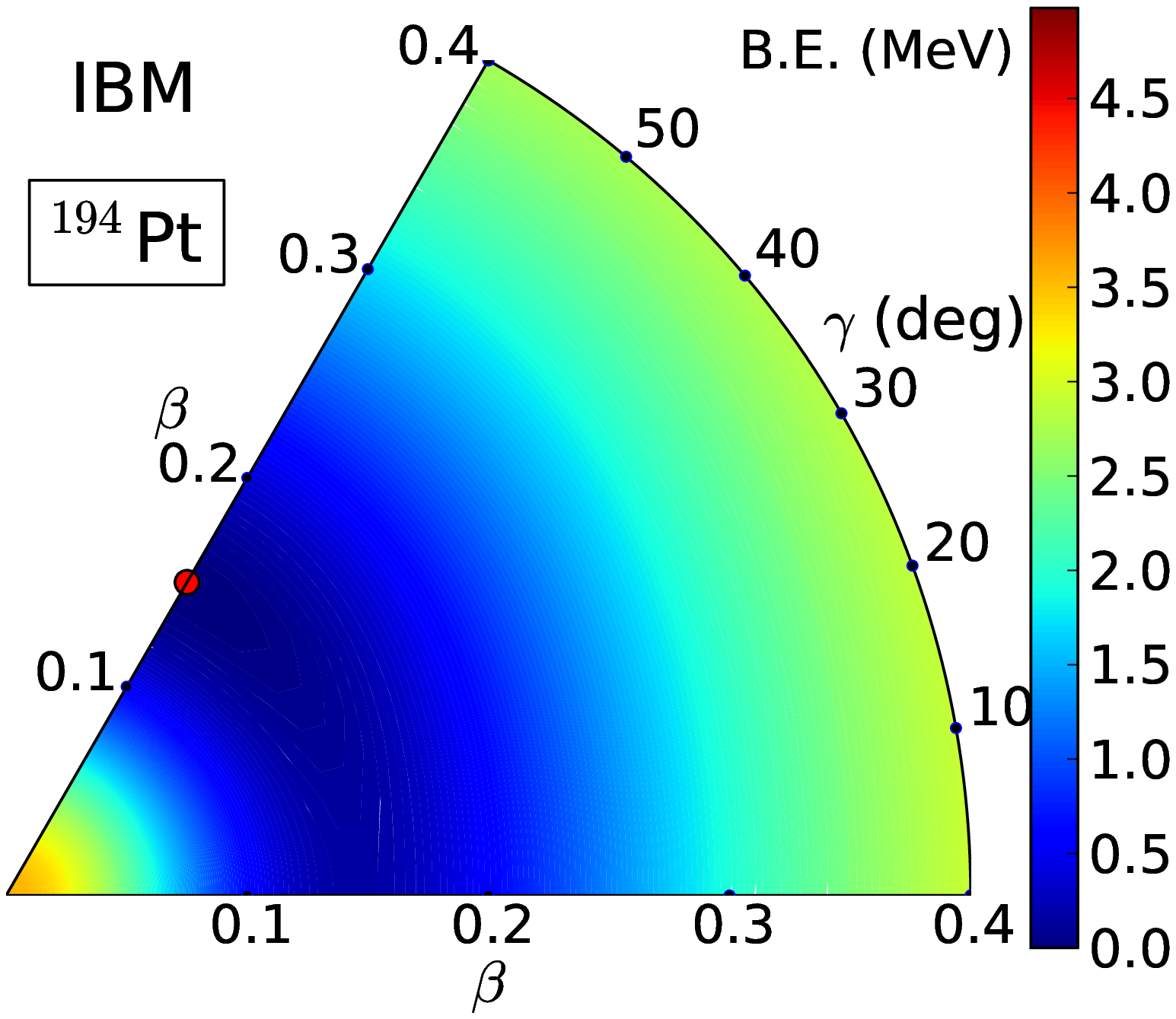}\\
\includegraphics[width=7.0cm]{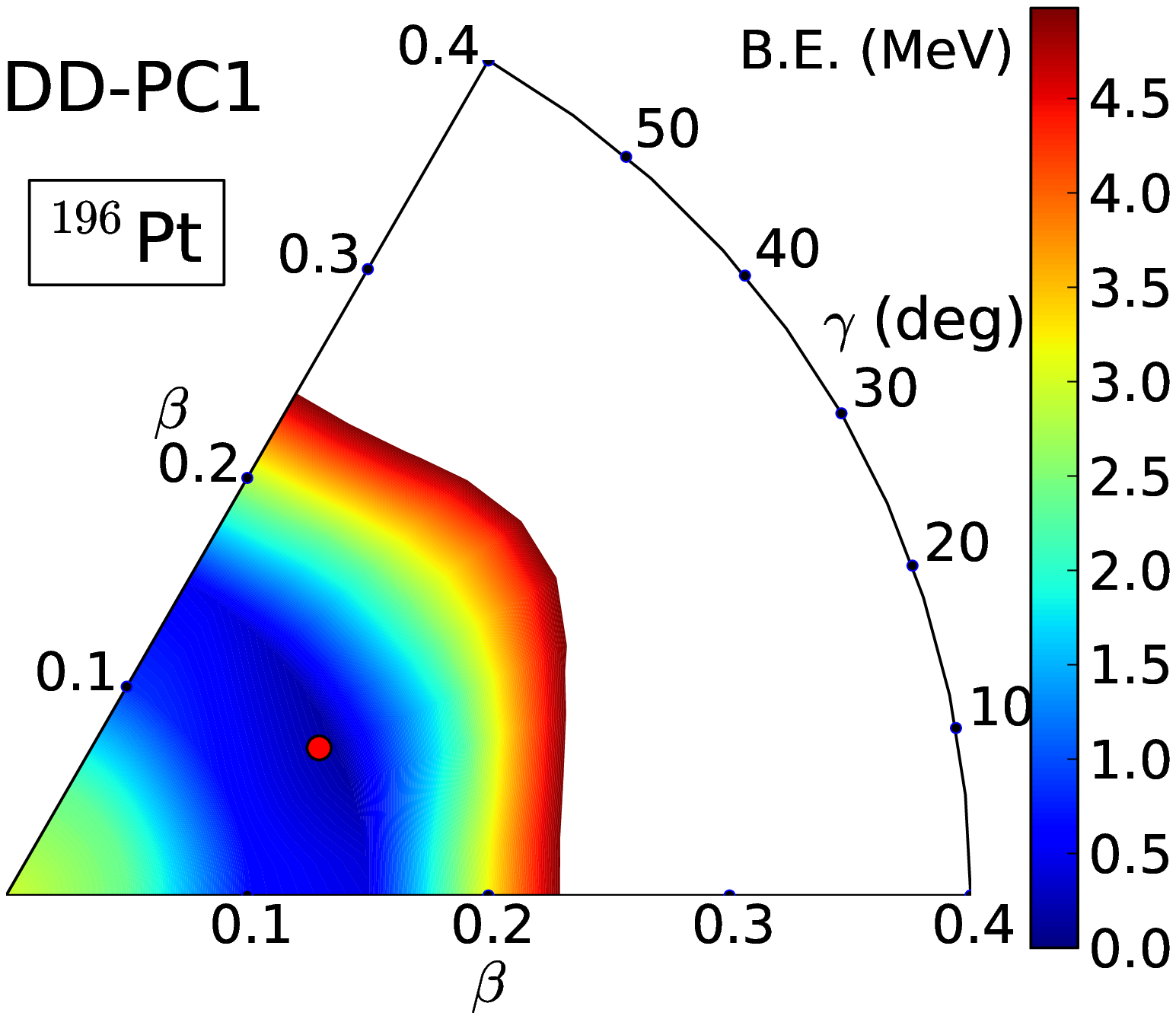} &
\includegraphics[width=7.0cm]{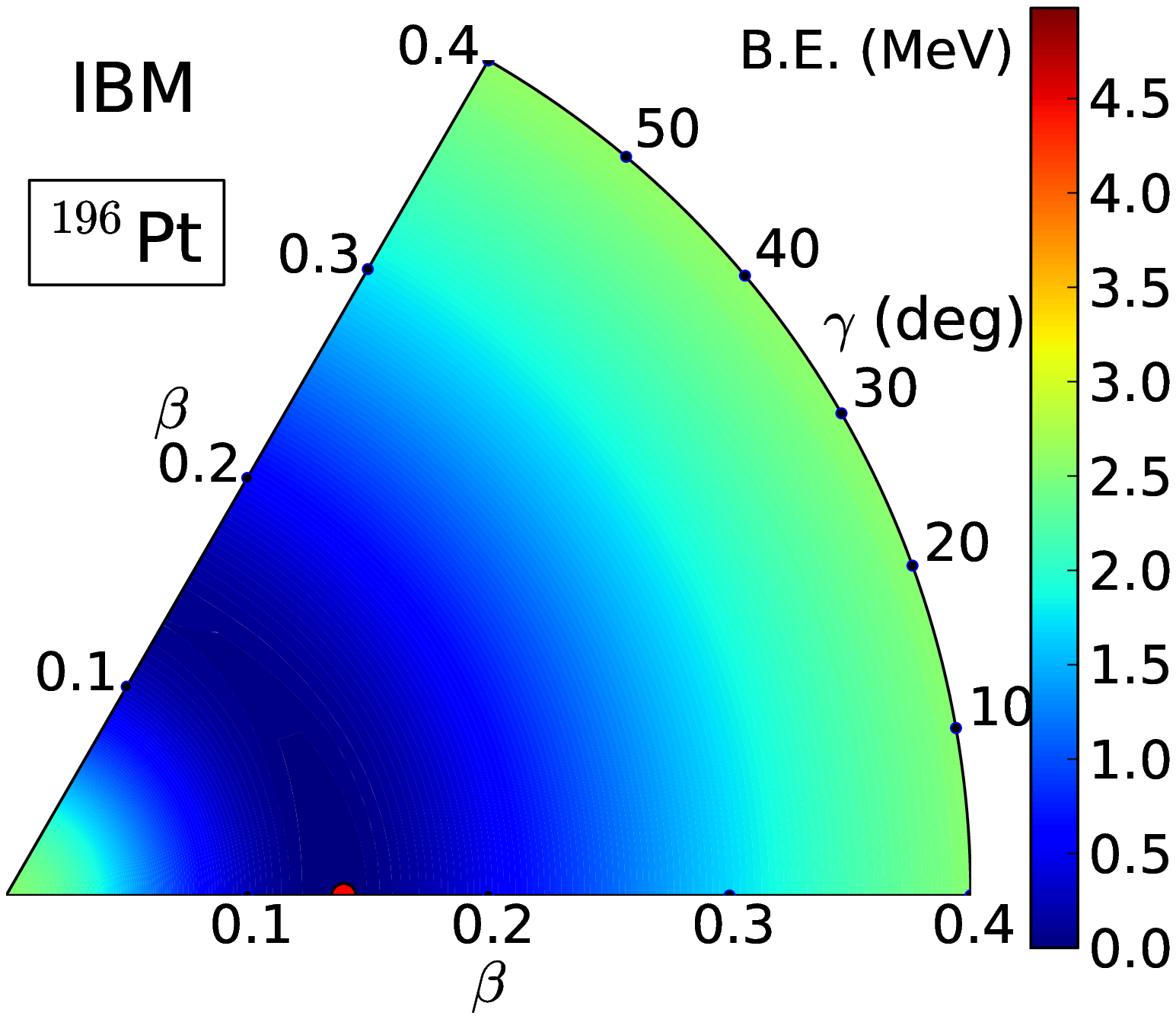}
\end{tabular}
\caption{(Color online) 
Self-consistent binding-energy maps of $^{192,194,196}$Pt  in the 
$\beta - \gamma$ plane ($0^{\circ} \leq \gamma \leq 60^{\circ}$), calculated with the 
RHB model using the DD-PC1 functional (left panels), and the corresponding
mapped energy surface of the IBM, $E_{\rm IBM}(\beta_{B},\gamma_{B})$.
 The IBM total energies are depicted in terms of $\beta$ and $\gamma$,
 where $\beta\propto\beta_{B}$ and $\gamma=\gamma_{B}$ (see text for definition). }
\label{fig:pes}
\end{center}
\end{figure*}

Most deformed nuclei display axially-symmetric prolate ground-state shapes, but few areas of the 
nuclide chart are characterized by the occurrence of non-axial shapes. One example is the 
$A \approx$ 190 mass region, where
both prolate to oblate shape transitions, and even triaxial ground-state shapes have been predicted. 

The left-hand side of Fig.~\ref{fig:pes}
shows the self-consistent RHB quadrupole binding energy maps of the $^{192,194,196}$Pt 
isotopes in the  $\beta - \gamma$
plane, calculated with the DD-PC1 energy density functional. 
The energy surfaces are $\gamma$-soft, with shallow
minima at $\gamma \approx 30^{\circ}$. In general the equilibrium deformation decreases 
with mass number and, proceeding to
even heavier isotopes, one finds that the energy map of $^{198}$Pt has also a non-axial minimum, 
whereas $^{200}$Pt displays a slightly oblate minimum \cite{RMF_review},
signaling the shell-closure at the neutron  
number $N=126$. On the right-hand side of Fig.~\ref{fig:pes}, we plot 
the corresponding IBM energy surfaces
$E_{\rm IBM}(\beta,\gamma)$, obtained by mapping each point of 
surface $E_{\rm RHB}(\beta,\gamma)$  
onto the energy surface calculated in
the IBM, following the procedure of Ref.~\cite{nsofull}. 
To be able to compare the low-energy spectra in the two models, 
the IBM surfaces are mapped in such a way to reproduce the RHB 
energy surfaces up to  $\approx 2$ MeV 
above the mean-field minimum. This means that the maps shown in 
Fig.~\ref{fig:pes} can only be compared for values of $\beta$ not very 
different from the minimum $\beta_{\rm min}$. For larger values of $\beta$, 
that is, for higher excitation energies the topology of the RHB surfaces is 
determined by single-nucleon configurations that are not included in the 
model space (valence space) from which the IBM bosons are constructed. 
For large  $\beta$-deformations, therefore, one  should not try to map the 
microscopic energy surfaces onto the IBM. This is the reason why the  
IBM energy surfaces are by construction always rather flat in the region
$\beta\gg\beta_{\rm min}$. 
In the vicinity of the minima the curvatures of the IBM energy 
maps are rather similar to those of the original RHB surfaces both in
$\beta$ and $\gamma$ directions. 
The derived values for the $\chi_{\pi}$ and $\chi_{\nu}$ parameters in
Eq.(\ref{eq:bh}) satisfy $\chi_{\pi}+\chi_{\nu}\sim 0$, characteristic for 
a $\gamma$-soft energy surface. 

One might notice that the IBM energy maps reproduce the 
value of $\beta$ at the minima predicted by the RHB calculation, whereas 
the mapping does not reproduce the shallow triaxial minima of the RHB surfaces. 
The minima of the IBM maps are either oblate or 
prolate. This is because the IBM Hamiltonian of Eq.~(\ref{eq:bh}) is too restricted 
to produce a triaxial minimum. 
In the analytical expression for $E_{\rm IBM}(\beta,\gamma)$ the 
$\gamma$-dependent term is proportional to $(\chi_{\pi}+\chi_{\nu})\cos{3\gamma}$,
and this places the minimum either on the prolate or oblate side according to
the sign of $(\chi_{\pi}+\chi_{\nu})$. 
The Pt nuclei considered here do not display any rapid structural change
but remain $\gamma$-soft. This feature appears to be independent of the choice 
of the EDF. A recent microscopic calculation using the Gogny-D1S EDF \cite{D1S}
also yielded shallow triaxial shapes, rather flat in the oblate 
region \cite{GognyIBMPt}, but quantitatively consistent with the present
analysis.  A similar trend was reported in other EDF-based studies of 
ground-state shapes of Pt isotopes \cite{RMFPt,gradient-2,RaynerPt}. 
In the present calculation the RHB surfaces become softer in $\gamma$ 
with increasing neutron number, and the softest nucleus is $^{196}$Pt. 
The corresponding IBM energy surfaces follow this evolution, but do not reproduce 
the triaxial minima because of the reasons explained above. 
The recent Gogny-EDF calculation \cite{GognyIBMPt}
predicts $^{192}$Pt to be the softest Pt isotope in this mass region. 

\section{Spectroscopic properties}
\label{sec:spectra}

\begin{figure}[ctb!]
\begin{center}
\includegraphics[width=8.5cm]{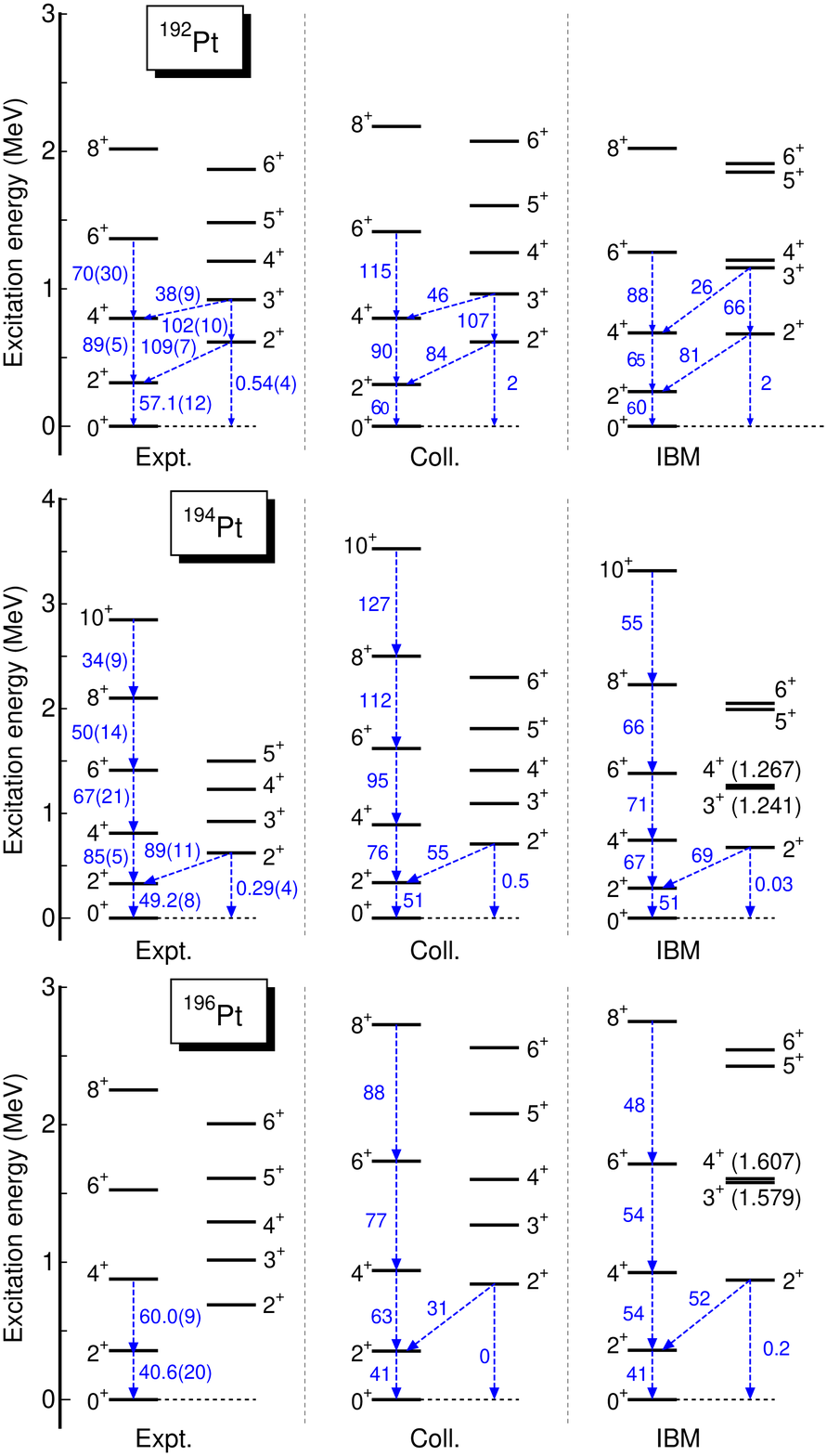}
\caption{(Color online) Low-lying collective spectra of $^{192,194,196}$Pt nuclei, 
calculated with the collective Hamiltonian based on the DD-PC1 functional and the 
corresponding IBM Hamiltonian, in comparison with available data. For each nucleus,
the $B$(E2) values (in Weisskopf units) obtained in the IBM are normalized 
to the $B({\rm E2};2^{+}_{1}\rightarrow 0^{+}_{1})$ predicted by the collective
Hamiltonian. The experimental excitation spectra and $B$(E2) values are from
Refs.~\cite{data} and \cite{BE2}, respectively. }
\label{fig:spectra}
\end{center}
\end{figure}

In Fig.~\ref{fig:spectra} we display the corresponding low-energy collective spectra
of $^{192,194,196}$Pt obtained from the collective Hamiltonian (middle panels), and the 
IBM Hamiltonian (panels on the right). The calculated ground-state
and (quasi) $\gamma$-vibration bands are compared to the corresponding sequences of experimental 
states \cite{data}. The eigenstates of the collective Hamiltonian in Eq.~(\ref{hamiltonian-quant})  are completely
determined by the DD-PC1 energy density functional plus a separable pairing interaction, and the transition
probabilities are calculated in the full configuration space using the bare value of the proton charge. 
Since $\hat H_{\rm IBM}$ in Eq.~(\ref{eq:bh}) acts only in the boson valence space, to calculate the $B$(E2) values 
one needs two additional parameters: the proton-boson and neutron-boson effective charges. For 
simplicity, here we take these effective charges to be equal, and in each nucleus normalize the $B$(E2) values 
obtained in the IBM to reproduce the transition probability $B({\rm E2};2^{+}_{1}\rightarrow 0^{+}_{1})$ 
calculated with the collective Hamiltonian. Thus we can only compare the ratios of the IBM $B$(E2) values, 
divided by $B({\rm E2};2^{+}_{1}\rightarrow 0^{+}_{1})$, to those predicted by the collective Hamiltonian 
based on DD-PC1, and to available data.

For the ground-state band, both the collective Hamiltonian and the IBM
predict excitation spectra in close agreement with experiment. 
For $^{192}$Pt, in particular, the calculated ground-state bands 
seem to indicate a somewhat larger deformation than observed experimentally.  
In fact, the theoretical energy ratio $R_{4/2}=E(4^{+}_{1})/E(2^{+}_{1})$ is 2.59
with collective Hamiltonian, and is 2.69 with the IBM Hamiltonian, compared to the experimental 
value $R_{4/2}=2.48$. A similar trend is also found for the other two nuclei.  
A more pronounced difference between the predictions of the two models 
is found in the E2 decay pattern of the ground-state band, particularly
in $^{194}$Pt nucleus for which data are available up to angular momentum 
$10^+$. For the spectrum calculated with the collective model, the 
E2 transition rates from the state with angular momentum
$L$ ($L\geqslant 2$) to the one with $L-2$ keep increasing as 
function of $L$, even though the corresponding experimental
$B$(E2) values in $^{192,194}$Pt decrease starting from $L=6$. 
The trend of the $B$(E2) values calculated with the IBM, on the other 
hand, is much closer to experiment.
The $B$(E2)'s decrease in the IBM because the model space is built 
from valence nucleons only, and the wave functions of higher angular-momentum 
states correspond to simple configurations of fully aligned $d$-bosons \cite{IBM}, 
whereas there is no limit on the angular momentum of eigenstates of the collective Hamiltonian. 

A more significant difference between the spectroscopic properties 
predicted by the collective Hamiltonian and the IBM is found 
in the sequence of levels built on the state $2^+_2$ -- the (quasi) $\gamma$-band. 
The IBM spectra display a staggering of excitation energies above $2^{+}_{\gamma}$,
with the formation of doublets ($3^{+}_{\gamma}$ $4^{+}_{\gamma}$), ($5^{+}_{\gamma}$
$6^{+}_{\gamma}$), ... etc, whereas the collective Hamiltonian yields 
a regular excitation pattern consistent with the experimental band. 
To be more precise, the IBM spectra correspond to 
$\gamma$-unstable nuclei, and are close to the limit of O(6) 
dynamical symmetry in which eigenstates of a boson Hamiltonian with the same 
$\tau$ quantum number are degenerate \cite{Pt196O6}. 
On the other hand, the $\gamma$-bands predicted by the collective model, 
as well as the experimental sequence, seem to be closer to rigid triaxiality \cite{triaxial}. 
The difference between the collective Hamiltonian and the IBM arises
probably because the shallow triaxial minima of the RHB
energy surfaces are not reproduced by  
the mapping onto the IBM total energy (cf. Fig.~\ref{fig:pes}).
The agreement of the IBM (quasi) $\gamma$-band with experiment could be improved 
by introducing additional interaction terms
in the IBM Hamiltonian, i.e., three-body terms (the so-called cubic
terms) \cite{cubic,Casten-cubic}. Terms of this type will have to be included 
for a more precise analysis and comparison of states above the yrast with 
experimental results. 

A nice feature of the present calculation, particularly the one with the
IBM Hamiltonian, is that the predicted 
$B$(E2) values for the transition $2^{+}_{2}\rightarrow 2^{+}_{1}$
are comparable to or even larger than those corresponding to 
$4^{+}_{1}\rightarrow 2^{+}_{1}$. 
This result is consistent with the experimental trend, whereas in
the recent Gogny-based  EDF calculation of Ref.~\cite{GognyIBMPt},
the $2^{+}_{2}\rightarrow 2^{+}_{1}$ transitions were much weaker than 
$4^{+}_{1}\rightarrow 2^{+}_{1}$. The corresponding Gogny energy surfaces 
displayed pronounced oblate minima in Ref.~\cite{GognyIBMPt}, unlike the present energy maps shown in
Fig.~\ref{fig:pes}.

\begin{figure*}[ctb!]
\begin{center}
\begin{tabular}{cc}
\includegraphics[width=7.0cm]{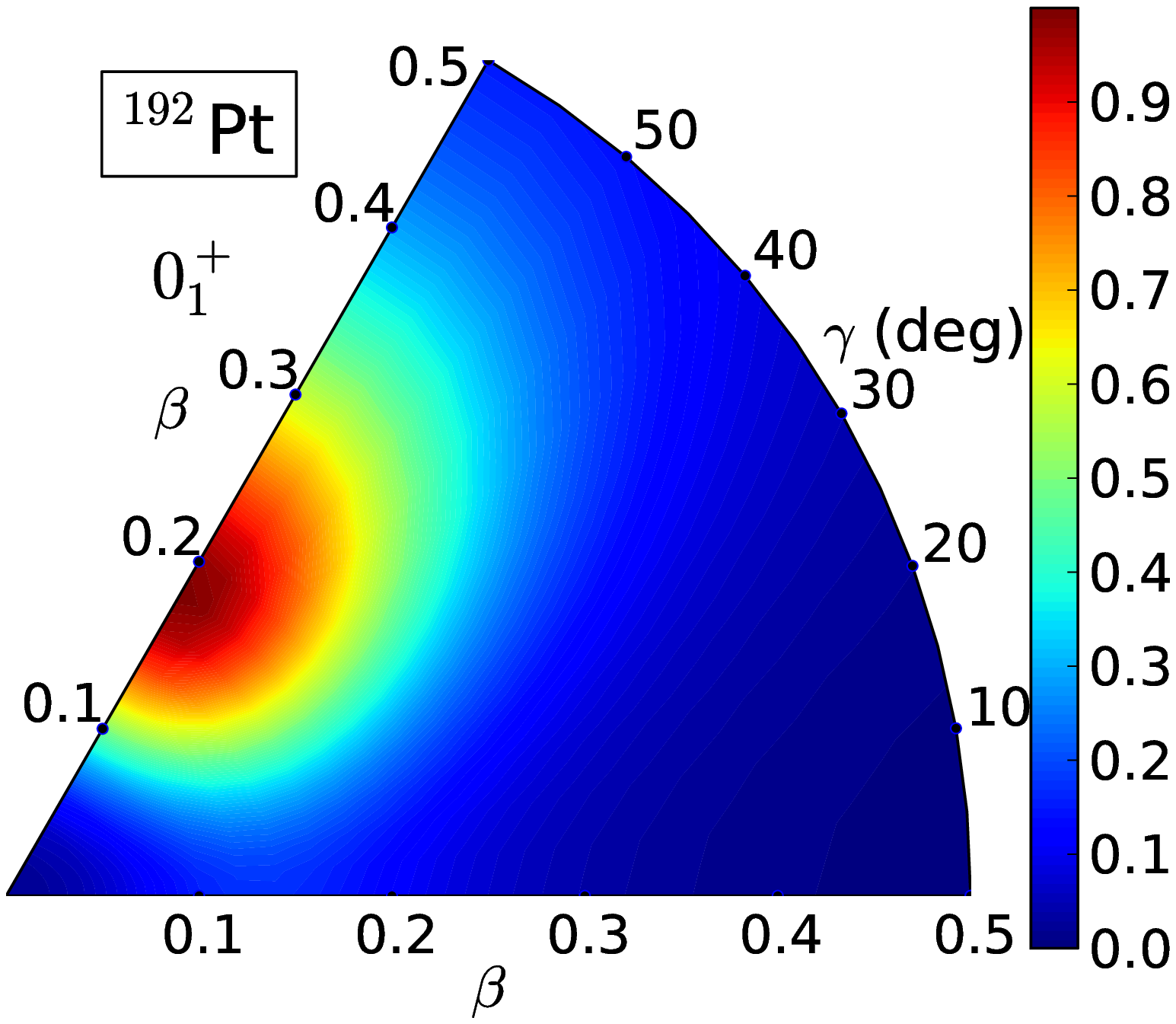} &
\includegraphics[width=7.0cm]{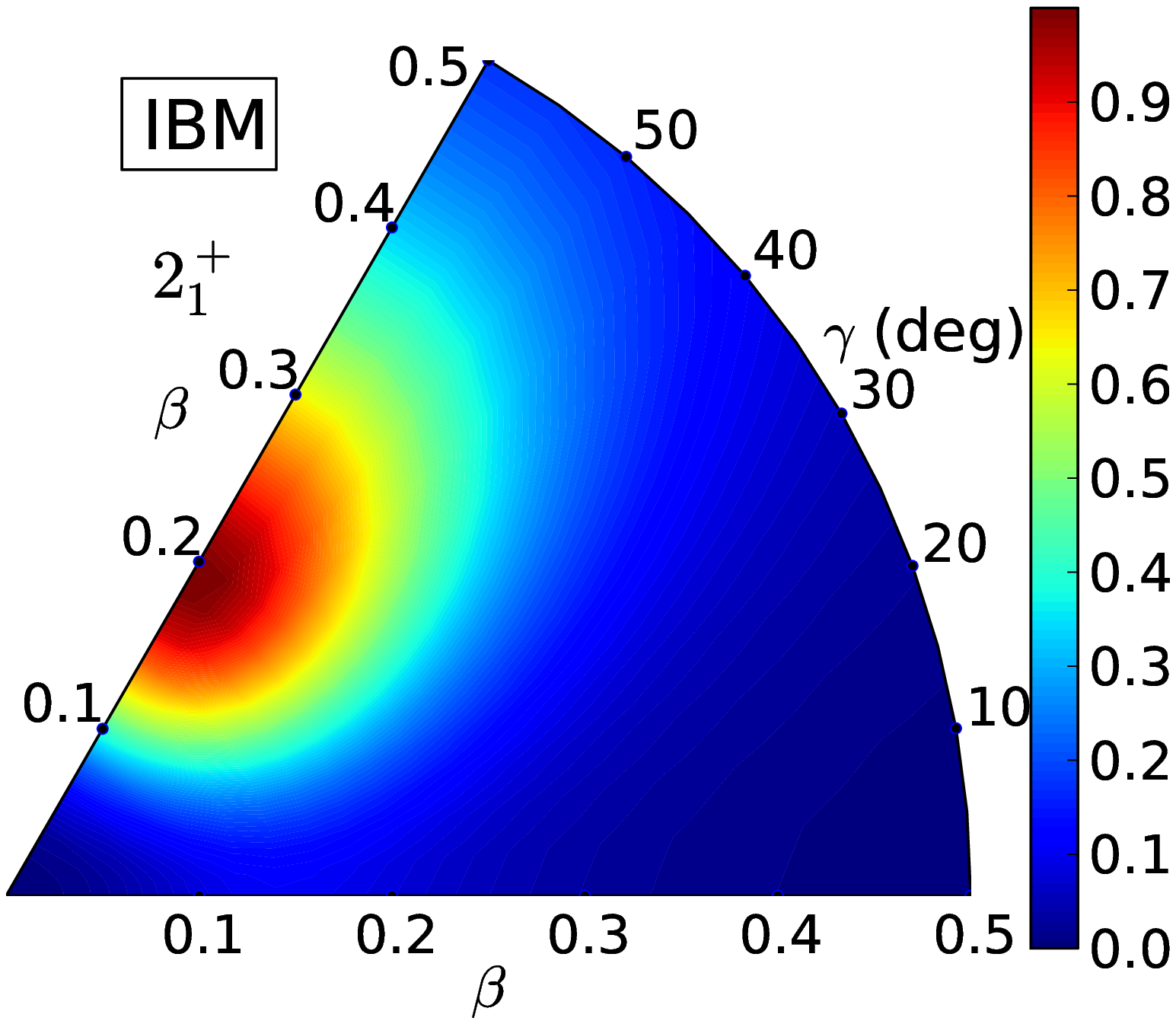}\\
\includegraphics[width=7.0cm]{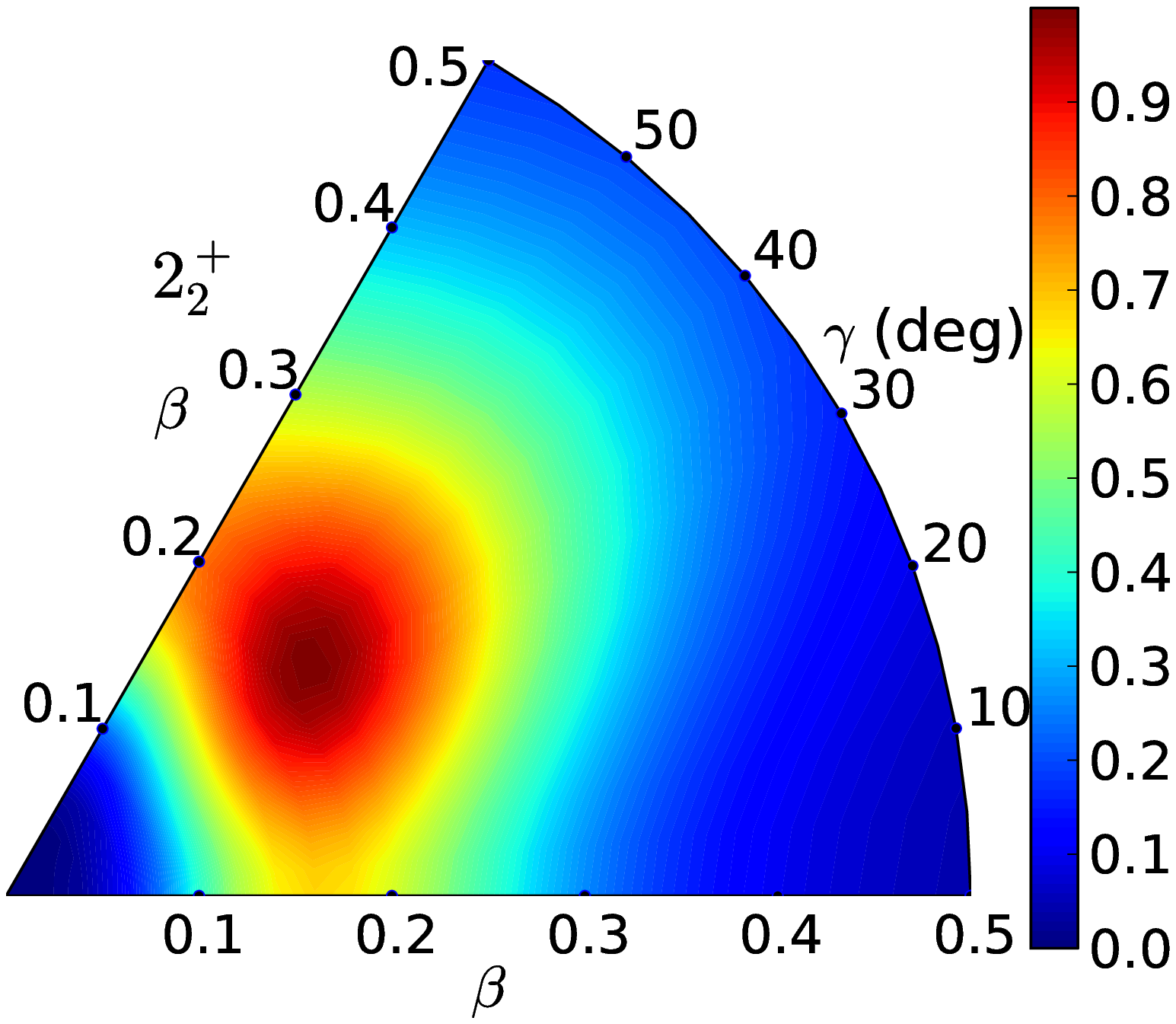} &
\includegraphics[width=7.0cm]{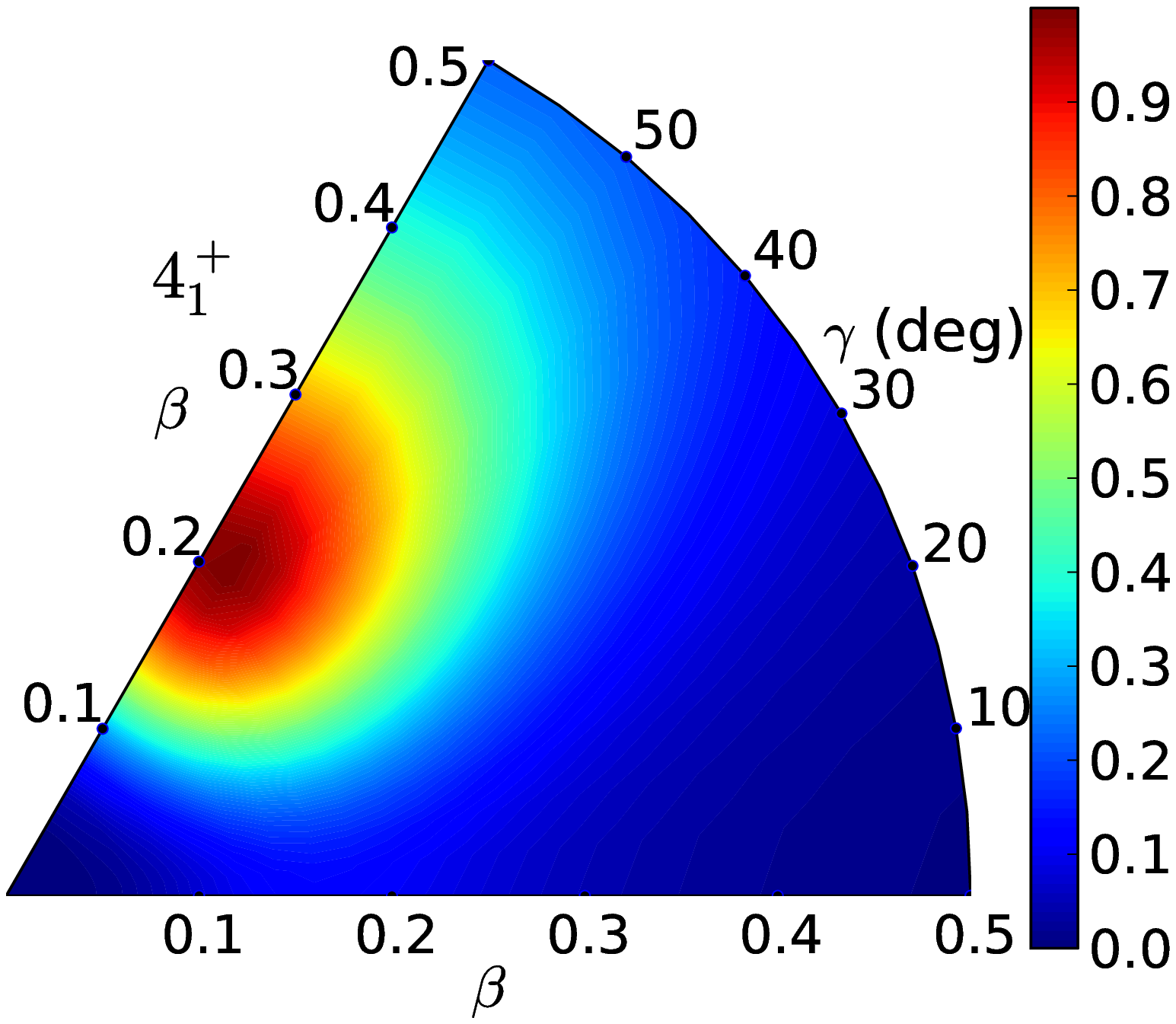}
\end{tabular}
\caption{(Color online) 
Absolute squares of the IBM wave functions in the $\beta - \gamma$ plane for the 
yrast states $0^{+}_{1}$, $2^{+}_{1}$, $4^{+}_{1}$, and the band-head of 
the $\gamma$-band $2^{+}_{2}$ of $^{192}$Pt. }
\label{fig:ovlp_ibm}
\end{center}
\end{figure*}

\begin{figure*}[ctb!]
\begin{center}
\begin{tabular}{cc}
\includegraphics[width=7.0cm]{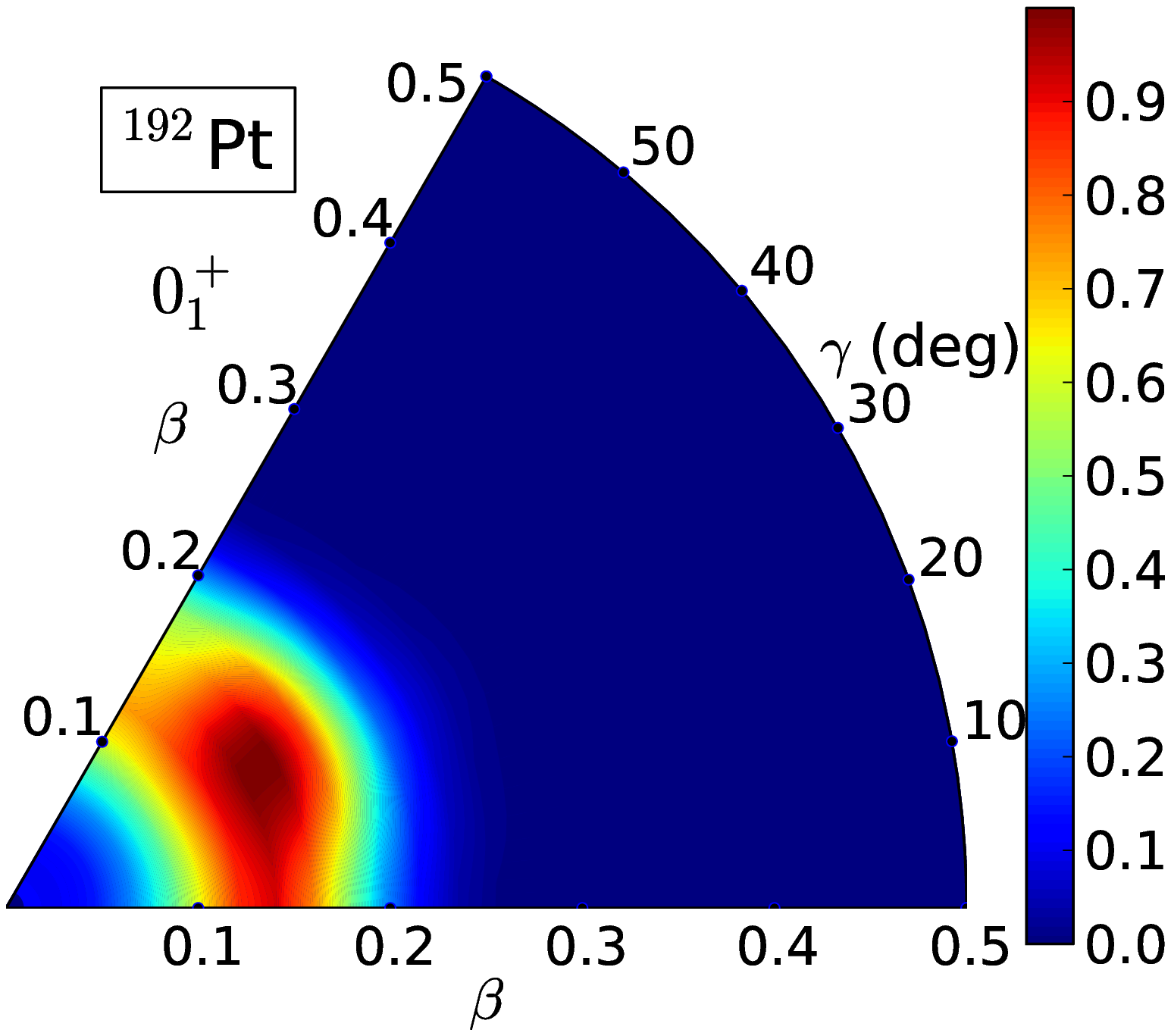} &
\includegraphics[width=7.0cm]{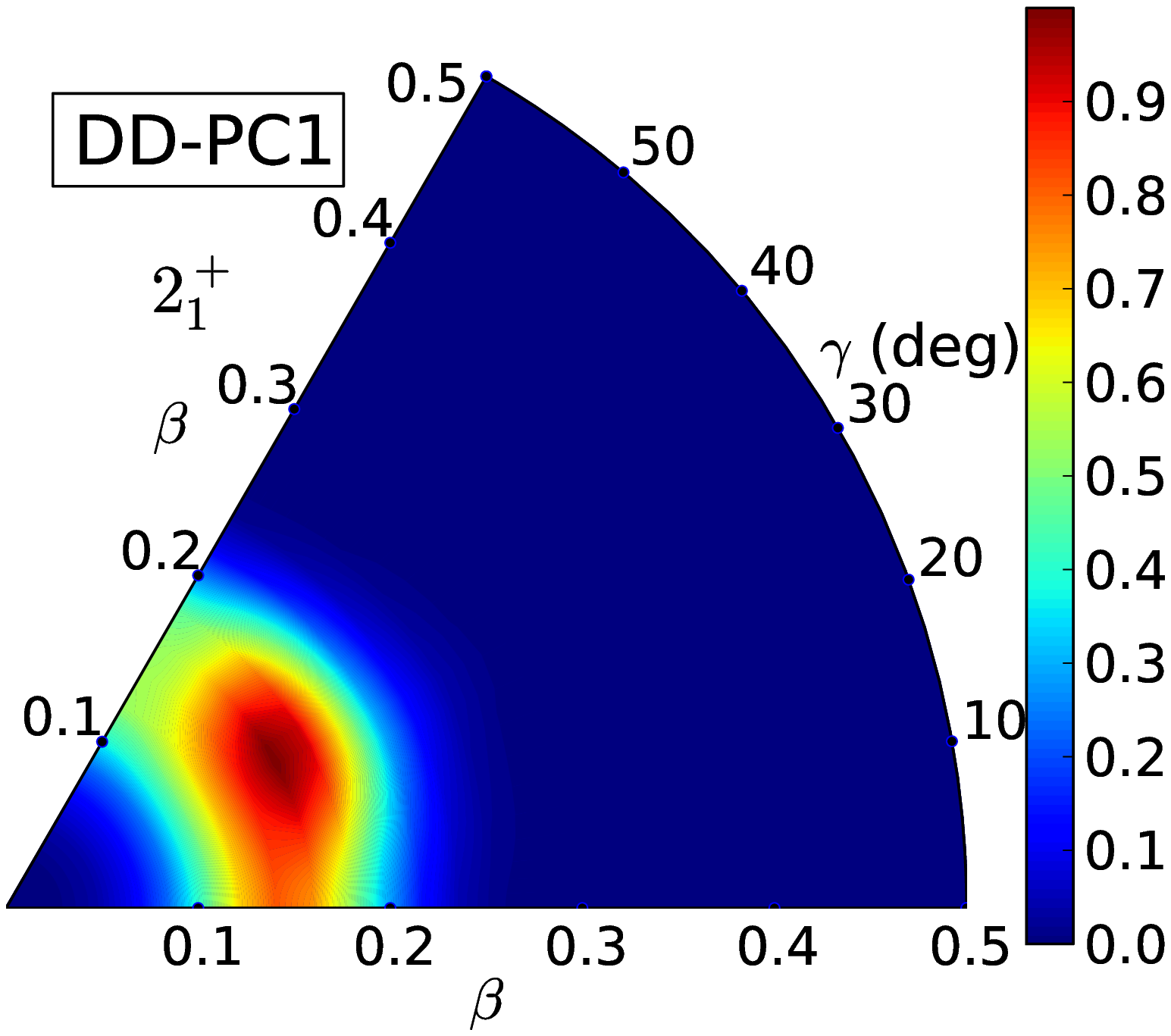}\\
\includegraphics[width=7.0cm]{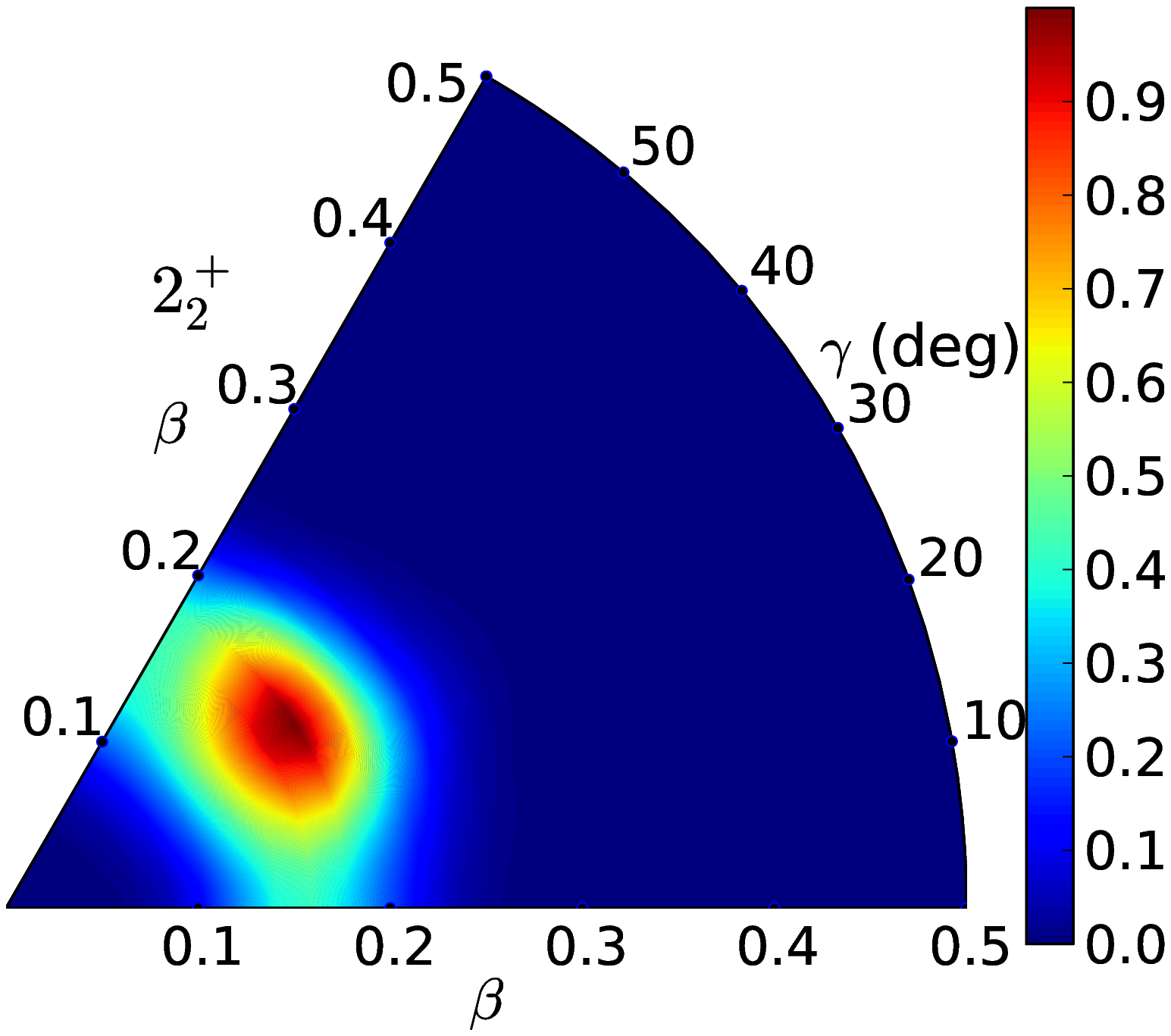} &
\includegraphics[width=7.0cm]{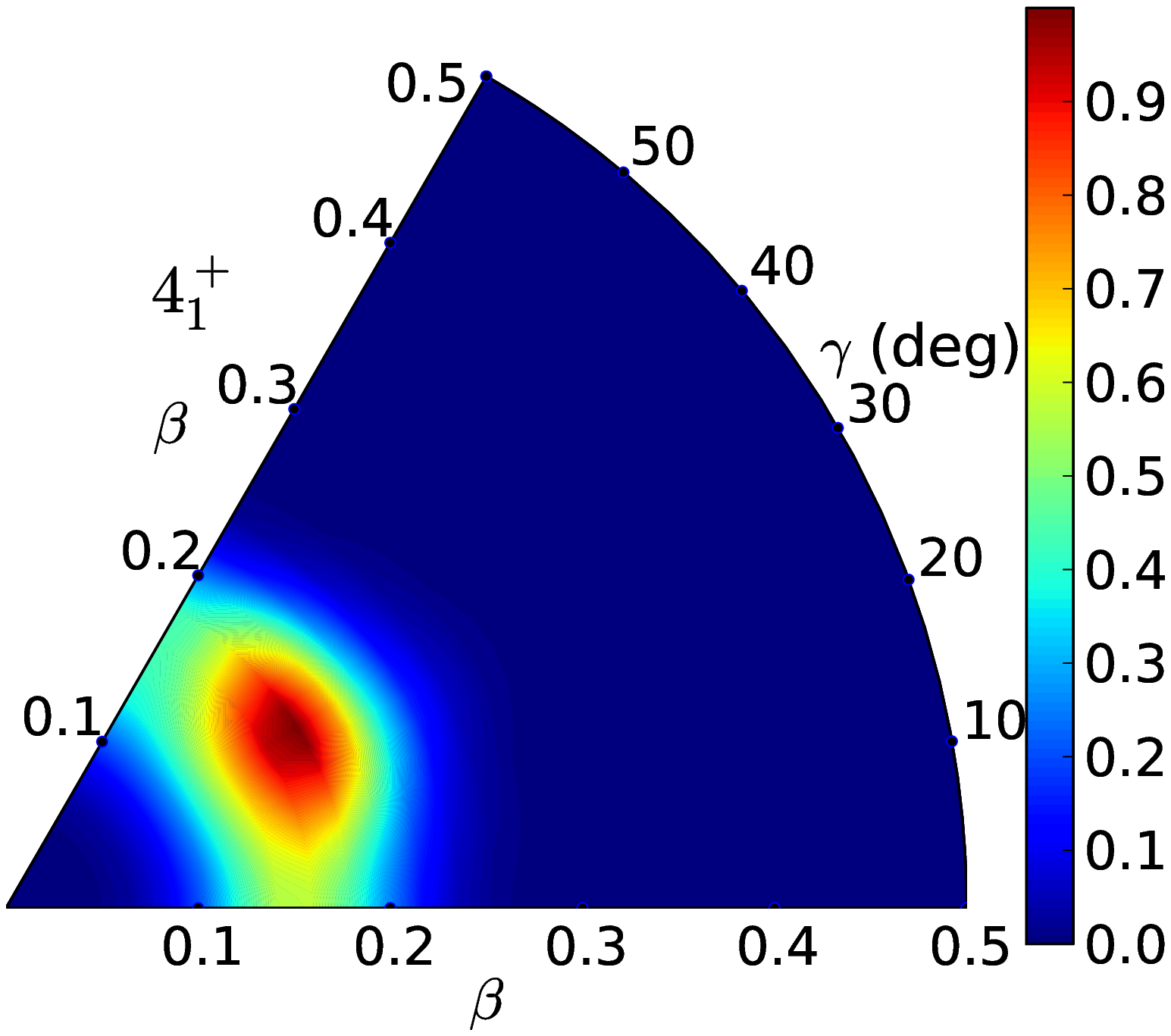}
\end{tabular}
\caption{(Color online) Same as described in the caption to Fig.~\ref{fig:ovlp_ibm}  but for the 
eigenstates of the collective Hamiltonian.}
\label{fig:ovlp_rmf}
\end{center}
\end{figure*}

Finally, in Figs.~\ref{fig:ovlp_ibm} and \ref{fig:ovlp_rmf} we compare the 
absolute squares of the collective wave functions for the 
yrast states $0^{+}_{1}$, $2^{+}_{1}$, $4^{+}_{1}$, and the band-head of 
the $\gamma$-band of $^{192}$Pt, calculated in the two models. These quantities are 
proportional to the probability density distributions in the $\beta - \gamma$ plane. 
Figure~\ref{fig:ovlp_ibm} shows the distributions  
$\sum_{M=-L}^{L}|\langle\Phi_{M}^{L}|\Psi(\beta,\gamma)\rangle|^{2}$, 
where $|\Phi_{M}^{L}\rangle$ denotes the IBM eigenstate for the state
with angular momentum $L$ and projection $M$.  
The wave functions of the yrast states are concentrated along the oblate axis, only for the 
state $4^{+}_{1}$ the maximum of the absolute square is located at $\gamma\sim 55^{\circ}$, 
and somewhat larger deviations from pure oblate configurations are found for higher 
angular momenta. For the state $2^{+}_{2}$, on the other
hand, the peak appears in the triaxial region ($\gamma\sim 35^{\circ}$), 
and the distribution is extended more toward oblate quadrupole 
deformations. The rather large overlap of the collective wave functions for the states 
 $2^{+}_{1}$ and $2^{+}_{2}$ explains the particularly strong 
$2^{+}_{2}\rightarrow 2^{+}_{1}$ transitions in this nucleus, and similarly in the 
other two Pt isotopes considered here. 
The corresponding absolute squares of the eigenstates of the collective Hamiltonian 
are shown in Fig.~\ref{fig:ovlp_rmf}. In this case already the wave functions of the 
yrast states reflect the $\gamma$-softness of the RHB energy surface, and 
the maxima of the absolute squares are found in the triaxial region of the 
$\beta - \gamma$ plane. 

\section{Conclusions and outlook}
\label{sec:summary}

Structure phenomena related to shape evolution currently present a very active research 
field in low-energy nuclear physics. Radioactive-beam facilities continue to provide interesting 
new data on shapes in regions of exotic nuclei far from stability. The
variation of ground-state shapes is, of course, governed by the evolution of the underlying 
shell structure of single-nucleon orbitals. It is, therefore, important to develop 
microscopic approaches that can be employed in quantitative analyses of shape phenomena and 
the resulting complex excitation spectra and decay patterns across the entire chart of nuclides. 
Such a framework is provided by nuclear energy density functionals (EDFs).

The advantages of EDFs are evident already at the basic level of implementation: 
an intuitive interpretation of self-consistent mean-field results in terms of intrinsic shapes 
and single-particle states, and the universality of EDFs that enables their
applications to all nuclei throughout the periodic chart. The latter is particularly 
important for extrapolations to regions of short-lived nuclei far from stability. 
When considering spectroscopic applications, the framework of EDF must be 
extended beyond the mean-field level to allow for a systematic treatment of
dynamical effects related to restoration of broken symmetries and fluctuations in collective 
coordinates. To calculate excitation spectra and transition rates, it is necessary to project 
states with good quantum numbers from the 
mean-field solution, and also take into account fluctuations
around the mean-field minimum. 

In this work we have compared two well known models that explicitly consider quadrupole collective 
correlations, both starting from maps of binding energy calculated with the same microscopic 
EDF. The first is the generalized collective Hamiltonian for quadrupole vibrations and rotations. 
The dynamics of the five-dimensional Hamiltonian is governed by the collective potential, the 
three vibrational mass parameters, and three moments of inertia for
rotations around the principal axes. These functions of the quadrupole deformation 
parameters are determined by constrained mean-field calculations using a given nuclear EDF. 
The diagonalization of the resulting Hamiltonian yields excitation energies and collective wave functions 
that can be used to calculate various observables. Calculations are performed in the full model space of 
occupied states (no distinction between core and valence nucleons, no need for effective charges). 
The second model considered in this work is the well-known and very successful IBM-2. 
In this approach the configuration space is first restricted to valence nucleons only, and further 
mapped to the space of $s$ and $d$ bosons. To determine the parameters of the IBM Hamiltonian, 
the energy surface calculated using a microscopic EDF, is mapped onto the corresponding 
boson energy surface under certain approximations. One then proceeds to calculate the excitation 
spectra and wave functions in the laboratory frame. To calculate transition probabilities, however, 
one needs to adjust the effective boson charges. Here this has been done so that in each 
nucleus the calculated $B({\rm E2};2^{+}_{1}\rightarrow 0^{+}_{1})$ coincides with the value 
obtained using the collective Hamiltonian. 

The two models have been compared here in a study of the evolution of non-axial shapes in 
Pt isotopes. Starting from the binding energy surfaces of $^{192,194,196}$Pt, calculated 
with the DD-PC1 energy density functional plus a separable pairing interaction, we have 
analyzed the resulting low-energy collective spectra obtained from the collective Hamiltonian, and 
the corresponding IBM-2 Hamiltonian.  The calculated ground-state
and $\gamma$-vibration bands have been also compared to the
corresponding sequences of experimental states. Both models predict that 
excitation energies and $B$(E2) values are in agreement with data. In particular, we notice
the excellent result for the predicted excitation energy of the band-head of the $\gamma$-band, 
as well as the good agreement with the experimental $B$(E2) values for transitions between the 
$\gamma$-band and the yrast band. 

There are also significant differences in the predictions of the two models. With the present form 
of the IBM Hamiltonian, restricted to two-body boson interactions, its expectation value in the 
boson coherent state does not reproduce the shallow triaxial minima of the binding energy 
maps predicted by the constrained self-consistent mean-field calculation
using DD-PC1. 
Since the mapped IBM energy surface is $\gamma$-soft rather than triaxial, the resulting spectra display a 
staggering of excitation energies above $2^{+}_{\gamma}$,
with the formation of doublets ($3^{+}_{\gamma}$ $4^{+}_{\gamma}$), ($5^{+}_{\gamma}$
$6^{+}_{\gamma}$), ... etc, in contrast to the regular excitation
pattern observed in experiment and reproduced by  the collective Hamiltonian. 
This problem could be solved by including three-body boson terms in
the IBM Hamiltonian, and work along this line is already in progress. 
When considering the calculated $B$(E2) values 
for transitions in the ground-state band, the IBM reproduces the gradual decrease of transition 
rates with angular momentum for $L \geq 6$, reflecting the finiteness of the valence space. 
On the other hand, even though the collective Hamiltonian predicts parameter-free $B$(E2) 
values in excellent agreement with experiment for transitions between low-spin states, the
calculated transition probabilities keep increasing with angular momentum, in contrast to data. 

Both models are based on binding energy surfaces calculated at zero rotational frequency. 
In general this leads to effective rotational moments of inertia that are lower than empirical 
values, that is, the calculated rotational bands are stretched in energy compared to experimental 
bands. In the collective Hamiltonian the moments of inertia can be improved by including the 
Thouless-Valatin dynamical rearrangement contributions. For the IBM Hamiltonian one needs 
to include the kinetic rotational term \cite{IBMrot}, and perform the mapping of microscopic energy surfaces 
calculated at finite values of the rotational frequency. We have already started with the 
implementation of these modifications in our current version of the collective Hamiltonian 
based on relativistic EDF, and in the IBM Hamiltonian. 
The comparison of the 
improved models will be the subject of a future study. 

\section*{Acknowledgment}

This work has been supported in part by Grants-in-Aid for Scientific
Research (A) 20244022 and No.~217368, and by MZOS - project 1191005-1010.
K.N. and D.V. acknowledge support by the JSPS. 
T.N. acknowledges support by the Croatian Science Foundation.

\end{document}